\documentclass[aps,prl,showpacs,floatfix,twocolumn]{revtex4}

\usepackage{bm}				
\usepackage{amsmath}
\usepackage{amssymb}
\usepackage{latexsym}
\usepackage{amsfonts}

\usepackage{graphicx}

\newcommand{\ket}[1]{\vert#1\rangle} 
\newcommand{\bra}[1]{\langle#1\vert} 
 
\newcommand{\proj}[1]{\ket{#1}\bra{#1}}

\newcommand{\identity}{\openone}

\newcommand{\abs}[1]{\left|#1\right|}

\begin{document}

\title{Entanglement evolution in finite dimensions}

\author{Markus Tiersch}
\email{markus.tiersch@physik.uni-freiburg.de}
\author{Fernando de Melo}
\author{Andreas Buchleitner}

\affiliation{Physikalisches Institut, Albert--Ludwigs--Universit\"at Freiburg,
Hermann--Herder--Str.~3, D--79104 Freiburg, Germany}

\date{\today}


\begin{abstract}
We provide a relation which describes how the entanglement of two $d$-level systems evolves as either system undergoes 
an arbitrary physical process. The dynamics of the entanglement turns out to be of a simple form, and is fully captured by a single quantity.
\end{abstract}

\maketitle

When precisely studying and manipulating the quantum world, the objects of interest such as photons~\cite{ref:haroche}, atoms~\cite{ref:meschede}, ions~\cite{ref:blatt} or quantum dots~\cite{ref:nakamura} are usually only few in number. But quantum mechanics \emph{does} allow also meso- or macroscopic systems to exhibit genuine quantum features such as interference~\cite{ref:interferenceOfBuckyBalls} and classically unachievable correlations~\cite{ref:schroedinger} -- the latter often lumped together under the label ``entanglement''. 
The description of these larger quantum systems, such as Bose--Einstein condensates, which consist of thousands of quantum objects, requires to resort to \emph{effective} properties since the quantum system's exact state surpasses what can analytically or numerically be coped with. 
Therefore, we need efficient theoretical and experimental tools to characterize and probe these quantum properties in terms of few and robust quantities~\cite{ref:walborn,ref:aolita,ref:kieselWeinfurter}.

Furthermore, since the transition from microscopic to macroscopic scales entails a rapidly increasing density of states, and, thus, a strongly enhanced fragility of generic quantum features to perturbations, we have to account for decoherence, or rather for the time scales on which quantum properties can prevail, in the presence thereof. Thus, we seek an efficient \emph{dynamical} description of the entanglement of \emph{open} quantum systems~\cite{ref:eberly,ref:mintertPhysRep,ref:unravelling}.

However, while sytematic progress has been achieved in describing the interference of ever larger quantum objects~\cite{ref:interferenceOfBuckyBalls}, in ever more complex environments, our systematic understanding of entanglement dynamics is still in its infancy. This lack of understanding is rooted in entanglement being a \emph{nonlinear} function of the system's density matrix. Thus, coherence is a necessary though by no means sufficient criterium for entanglement to prevail. Consequently, the only systematic results on entanglement dynamics were hitherto available for the smallest -- microscopic -- quantum systems which can harbour entanglement, i.e.\ for pairs of qubits~\cite{ref:evolEqua2x2}. Here, we present a first general result which describes the entanglement dynamics of two-party quantum systems with arbitrary (finite) dimensions of their components.

The setup we consider consists of two initially entangled $d$-level systems, of which one undergoes an arbitrary physical process, in general some open system dynamics in which it interacts with uncontrolled and not measurable degrees of freedom of its environment. Such processes are often referred to as channels, maps or superoperators~\cite{ref:Preskill,ref:Kraus,ref:Nielsen}. We denote them by $\$$. Starting with a pure state $\ket{\chi}$ the system's final state then takes the form
\begin{equation} \label{eq:initial}
\rho^\prime = (\identity \otimes \$) \proj{\chi}
\,,
\end{equation}
which now needs to be characterized in terms of its entanglement content. As compared to the simple case $d=2$ in \cite{ref:evolEqua2x2}, the following complication arrises: $\rho^\prime$ may be a mixture of pure states $\ket{\psi_i}$ which are living on \emph{different}, \emph{strict subspaces} of the $d$-dimensional space, i.e.\ the Schmidt rank of $\rho^\prime$ may drop below $d$, but this no longer  implies separability.
We will here focus on the entanglement evolution on time scales while $\rho^\prime$ preserves its initial Schmidt rank, though will see that the infered dynamics has some bearing also for the entanglement evolution on longer time scales.

We quantify the entanglement exhibited by $\rho^\prime$ using G-concurrence~\cite{ref:Gconc}, which reduces to concurrence~\cite{ref:conc} when restricting to two two-level systems. For a pure state $\ket{\chi}$ it is the geometric mean of its $d$ Schmidt coefficients $\lambda_i$ $(i=1,\dots,d)$. However, when $\ket{\chi}$ is given as $\ket{\chi}=\sum_{i,j=1}^{d} A_{ij} \ket{i}\ket{j}$ with basis states $\ket{i}$ and $\ket{j}$ for the respective subsystems, G-concurrence is more conveniently evaluated by
\begin{equation}
G_d (\ket{\chi}) = d \left[ \det (A^\dag A) \right]^{1/d}
\,.
\end{equation}
For mixed states $\rho$, G-concurrence is calculated through the usual minimization procedure of the ensemble's average,
$G_d(\rho) = \inf \sum_i p_i G_d(\ket{\phi_i})$,
over all possible decompositions into pure states, i.e.\ $\rho=\sum_i p_i \proj{\phi_i}$ (with $p_i>0$ and $\sum_i p_i=1$).
An attempt for an analytical result of this optimization procedure similar to the one done by Wootters for concurrence~\cite{ref:conc} yielded computable upper and lower bounds~\cite{ref:GconcMixed}.
The same framework also allows to consider $(d \times f)$-systems, $d \leq f$, since pure states (and the pure state vectors that form a decomposition of a mixed state) of such a system can exhibit entanglement in at most $d$~levels, by virtue of their Schmidt-representation.

Exploiting some specific algebraic properties of G-concurrence, we can largely follow the line of argument when describing entanglement dynamics of $2 \times 2$ states~\cite{ref:evolEqua2x2}.
In order to evaluate the entanglement of the final state $\rho^\prime$, we first need to express the initial pure state $\ket{\chi}$ as the result of a so-called filtering operation $M_\chi$ \cite{ref:GisinFiltering} acting on either party of a maximally entangled state $\ket{\Phi}$, i.e.\
\begin{equation}
\ket{\chi} = (M_\chi \otimes \identity) \ket{\Phi}
\,.
\end{equation}
$M_\chi=\sqrt{d} \sum_{i,j=1}^d A_{ij} \ket{i}\bra{j}$ here acts on the first subsystem of $\ket{\Phi}=\sum_{n=1}^d \ket{n}\ket{n}/\sqrt{d}$. Note that such filtering is always possible, for arbitrary $\ket{\chi}$, given its matrix representation $A_{ij}$. 
The channel represented by the filtering operation $M_\chi$ and the state $\ket{\chi}$ are isomorphic~\cite{ref:Jamiol}.

Given the filtering $M_\chi$ and the channel $\$$, which  act on different and possibly spatially separated parts of the system, the temporal order of their execution must not be of influence to the final state, and we can exchange their order in our representation of $\rho^\prime$,
\begin{equation} \label{eq:dual}
\rho^\prime = (M_\chi \otimes \identity) \rho_\$ (M_\chi^\dag \otimes \identity)
\,,
\end{equation}
where we introduced the result of the channel $\$$ acting on the maximally entangled state, $\rho_\$ = (\identity \otimes \$) \proj{\Phi}$, which is mixed in general. Again, state $\rho_\$$ and channel $\$$ are related to each other via the Jamio{\l}kowski isomorphism~\cite{ref:Jamiol}. Summarizing these two steps above, we have transformed the initial setup via the Jamio{\l}kowski isomorphism into the dual one, where the role of states  ($\rho_\$$ replaces $\ket{\chi}$) and channels ($M_\chi$ replaces $\$$) is interchanged. As in \cite{ref:evolEqua2x2} one may also arrive at this point by inserting an intermediate teleportation procedure before the $d$-level system undergoes the action of the channel.

Having arrived at this particular form \eqref{eq:dual} of the final state $\rho^\prime$, our chosen entanglement measure, G-concurrence for a $(d \times d)$-system, factorizes much as concurrence does for the $(2 \times 2)$-case~\cite{ref:evolEqua2x2}.
That is, G-concurrence exhibits the particular property that a single operator acting on either one of the subsystems simply factors out~\cite{ref:Gconc,ref:VerstraeteConcFact}:
$G_d [(M \otimes \identity) \ket{\psi}] = \abs{\det(M)}^{2/d} G_d(\ket{\psi})$. Applying this to the dual form of the final state \eqref{eq:dual} (in this very form~\cite{ref:noteOptimization}), and realizing that the determinant of $M_\chi$ relates to the G-concurrence of the initial state $\ket{\chi}$, yields our core result
\begin{equation} \label{eq:result}
G_d \big[ (\identity \otimes \$)\proj{\chi} \big]
=
G_d (\ket{\chi}) \;
G_d \big[ (\identity \otimes \$) \proj{\Phi} \big]
\,.
\end{equation}
The entanglement of two $d$-level systems in terms of G-concurrence evolves equally for all pure states $\ket{\chi}$, is solely given by the evolution of a maximally entangled state, and merely rescaled by the initial entanglement. This effectively reduces the vast space of initial conditions to a single one.

For general, i.e.\ mixed, initial states $\rho_0$ and/or two one-sided channels $\$_1 \otimes \$_2$, the convexity property of entanglement monotones~\cite{ref:Gconc,ref:BennettMixedStateEntang,ref:VidalEntanglementMonotones} leads to an inequality instead, which provides an upper bound
\begin{multline}
G_d \big[ (\$_1 \otimes \$_2) \rho_0 \big]
\leq
G_d (\rho_0) \times \\
\times
G_d \big[ (\$_1 \otimes \identity) \proj{\Phi} \big]
G_d \big[ (\identity \otimes \$_2) \proj{\Phi} \big]
\,.
\end{multline}
Here equality holds, for example, in case of pure initial states and two channels, if the filtering operation $M_\chi$ and either of the channels, $\$_1$ or $\$_2$, commute -- such that the order of the execution of the channels and of the filtering can be interchanged, in order to achieve a factorization as found above for one-sided channels.

Let us contemplate the physical content of eq.~(\ref{eq:result}): 
Since G-concurrence -- computed from the \emph{product} of \emph{all} Schmidt-coefficients of a state -- vanishes if at least one of the Schmidt coefficients is zero, it measures entanglement in exactly $d$ levels. Whenever the bipartite system under study undergoes some dynamics which induce a redistribution of amplitudes to strictly less than $d$ levels, G-concurrence vanishes, while entanglement on a strict subset of levels may prevail.  
For a channel with a continuous time evolution, e.g. governed by a Lindblad-type equation, the system's state evolves continuously within the set of states. 
Thus, when initially prepared in a state with non-vanishing G-concurrence, it is guaranteed to exhibit non-vanishing G-concurrence at least during an initial, \emph{finite} time interval, and the time evolution of entanglement during this period is described by our result~\eqref{eq:result}.
At later times, one has to rely on a hierarchy of entanglement monotones $C_k(\ket{\psi})$, $k<d$, for example those which were defined along with G-concurrence (being the last member of the hierarchy $G_d \equiv C_d$)~\cite{ref:Gconc}. Similarly to G-concurrence, they are computed from the sum of all different products of $k$ Schmidt coefficients, and hence capture the entanglement of exactly $k$ levels. For them we similarly derive
\begin{equation}
C_k \big[ (\identity \otimes \$) \proj{\chi} \big]
\leq
C_k (\ket{\chi}) \;
C_k \big[ (\identity \otimes \$) \proj{\Phi} \big] \;
\binom{d}{k}^{1/k}
\,,
\end{equation}
with an additional binomial coefficient~\cite{ref:noteMonotones} which accounts for the different possibilities to select $k$ out of $d$ levels.

Finally, we conclude with a remarkable observation which suggests that the time evolution of G-concurrence is intimately related to the evolution of the entanglement as quantified by concurrence~\cite{ref:rungtaConc}:
For this we choose an entangled pure state with non-vanishing G-concurrence and expose one of the $d$-level systems for a time $t$ to a depolarizing environment.
Such environments completely destroy the information about a system's state without any bias and thus turn it into a totally mixed state. When viewing the quantum system as an information carrier, this means that all possible independent errors occur with the same probability.
The dynamics of G-concurrence during this procedure is then, by virtue of~\eqref{eq:result}, entirely determined by the dynamics of the maximally entangled state. Depolarizing one subsystem thereof produces isotropic states~\cite{ref:HorodeckiIsoStates}, which can be parameterized by 
\begin{equation}
\rho_F = \frac{1-F}{d^2-1} \big( \identity - \proj{\Phi} \big) + F \proj{\Phi}
\,.
\end{equation}
The state evolves along the line segment parametrized by $F$ according to the solution of the Lindblad equation (Markovian environment assumed) with interaction rate $\Gamma$, what results in
\mbox{$F(t)=[1+(d^2-1)\exp(-2d\Gamma t)]/d^2$}. Thus, initially with $F(0)=1$ the maximally entangled state is recovered, and thereafter the state closes in towards the totally mixed state $\identity/d^2$, at $F=1/d^2$, asymptotically in time.
The condition that an isotropic state be of Schmidt number $k$~\cite{note:Schmidtnumber,ref:TerhalSchmidtNumb}, namely if and only if $k-1 < Fd \leq k$~\cite{ref:TerhalSchmidtNumb}, then determines the points in time when the Schmidt number drops. Thus, the drop from Schmidt number $k$ to $k-1$ occurs at time $t_k=\ln[(d^2-1)/(dk-d-1)]/(2d\Gamma)$ and hence entanglement measures quantifying only entanglement in $k$ levels vanish.
In particular, for $F(t_2)=1/d$ with $t_2=\ln(d+1)/(2d\Gamma)$, the state turns separable, whereas for \mbox{$F(t_d)=(d-1)/d$}, at time \mbox{$t_d=\ln[(d^2-1)/(d^2-d-1)]/(2d\Gamma)$}, G-concurrence disappears.
The inverse of this depletion time of G-concurrence then constitutes a characteristic decay rate $\Gamma_G$ compared to the decay rate $\Gamma_C$ of the generalized concurrence~$C$~\cite{ref:rungtaConc} (which vanishes if and only if the state is separable).
%
\begin{figure}
	\centering
	\includegraphics[width=\linewidth]{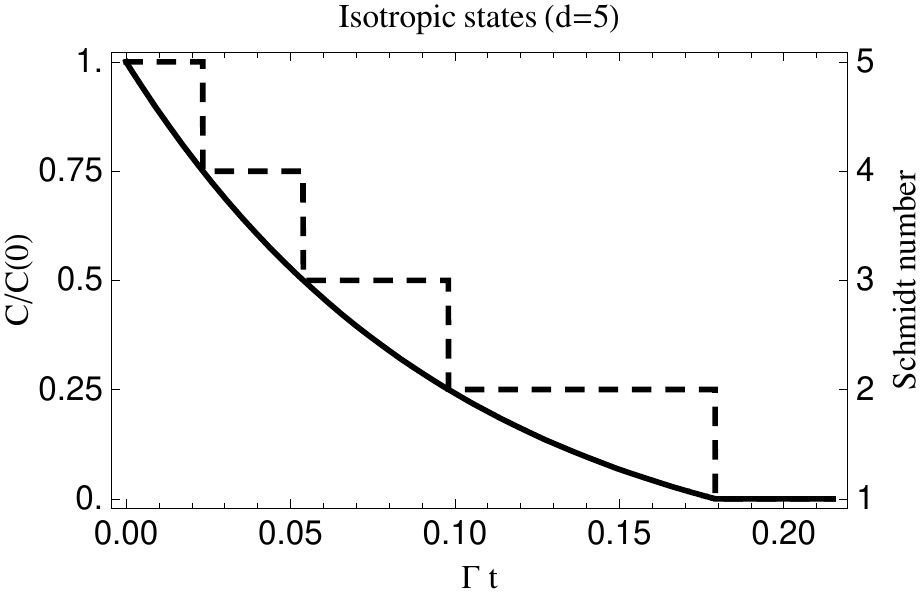}
	\caption{Evolution of concurrence (solid) and Schmidt number (dashed) as one party of the maximally entangled state $\ket{\Phi}=\sum_{n=1}^d\ket{n}\ket{n}/\sqrt{d}$, ($d=5$) passes through a depolarizing channel with rate $\Gamma$ (resulting in an isotropic state). The initial decay of entanglement is well approximated (linearly) by the entanglement at time 0 ($C(0)$) and when the Schmidt number drops for the first time (here from 5 to 4, with the initial entanglement reduced by a fraction $1/(d-1)$, from 1 to $3/4$), which is exactly when G-concurrence vanishes.}
	\label{fig:plot}
\end{figure}
%
Due to the high symmetry of isotropic states the infimum optimization for the concurrence of these high-dimensional mixed states can be carried out analytically~\cite{ref:rungtaConcIsotrop}, and, when normalized to the initial concurrence, yields
\begin{equation}
C(\rho_F) = (Fd-1)/(d-1)\, ,
\end{equation}
for $1/d \leq F \leq 1$, and $C(\rho_F)=0$ otherwise. For the chosen dynamics, that are encoded in $F(t)$, it decays exponentially with a small offset, that guarantees it to vanish at time $t_2$. Its time derivative then determines the rate $\Gamma_C (t)$ at which concurrence decays. Comparing the characteristic decay rate $\Gamma_G$ of G-concurrence to the decay rate $\Gamma_C (t)$ of concurrence, at short times as illustrated in Fig.~\ref{fig:plot}, yields a proportionality factor which for any initial pure state only depends on the system dimension:
\begin{equation}
\frac{\Gamma_G}{\Gamma_C (0)}
=
\frac{d}{(d+1) \ln \left(\frac{d^2-1}{d^2-d-1}\right) }
\stackrel{d \gg 1}{\approx}
d
\,.
\end{equation}
This indicates that the dynamics of G-concurrence and concurrence are much interrelated with one-another although G-concurrence detects only a very specific type of entanglement, that is entanglement in exactly $d$ levels. It also suggests that G-concurrence already exhibits most properties of the evolution of entanglement in general -- in particular at short times, when entanglement control is most crucial for possible applications.

We enjoyed illuminating discussions and exchanges with Thomas Konrad and Christian Kasztelan. Moreover, helpful comments by Marco Piani, Jens Eisert, and Fernando Brand\~ao are gratefully acknowledged.
F.~de M.~also acknowledges the financial support by the Alexander von Humboldt Foundation.




\begin{thebibliography}{499}
\bibitem{ref:noteOptimization}
The factorization itself occurs inside the optimization procedure for the mixed state $\rho_\$$:
The G-concurrence of the final state evaluates as
$G_d(\rho^\prime) =
\inf_{\{p_i,\ket{\phi_i}\}} \sum_i p_i G_d[(M_\chi\otimes\identity)\ket{\phi_i}]$,
where the $\{p_i,\ket{\phi_i}\}$ define a pure state decomposition of $\rho_\$$. It is sufficient to optimize over a pure state decomposition of $\rho_\$$ instead of $\rho^\prime$ (given by $\{p_i,(M_\chi\otimes\identity)\ket{\phi_i}\}$) if the $M_\chi$ are invertible -- which is the case if $\ket{\chi}$ is entangled in $d$-levels.
\bibitem{ref:noteMonotones}
We used the definition of $C_k$ in terms of the $k$-th compound matrix in \cite{ref:Gconc}. The resulting Hilbert--Schmidt norm is then factorized using the Cauchy--Schwarz inequality.
\bibitem{note:Schmidtnumber}
Mixed states with Schmidt number $k$ cannot be decomposed into pure states which all have Schmidt rank smaller than $k$;
formally: $k=\min_{\{p_i,\ket{\phi_i}\}}[\max_i k(\ket{\phi_i})]$.
\end{thebibliography}
\end{document}